\documentclass[letterpaper]{IEEEtran}

\usepackage{cite}                        
\usepackage[pdftex]{graphicx}                      
\usepackage[cmex10]{amsmath}      		  
\usepackage{array}                  
\usepackage{mdwmath}    
\usepackage{mdwtab}                         	
\usepackage{amssymb}
\usepackage{comment}
\usepackage[english]{babel} 
\usepackage{mathtools,amssymb,lipsum, nccmath}        	

\usepackage{url}

\begin{document}

\markboth{This article has been accepted for publication. DOI: 10.1109/ACCESS.2017.2786467, IEEE Access}{}

\title{
	\begin{center}
	Vortex Waves and Channel Capacity: \\
	Hopes and Reality
	\end{center}
	}

\author{R. Gaffoglio, A. Cagliero, G. Vecchi, {\it Fellow}, {\it IEEE}, F. P. Andriulli, {\it Senior Member}, {\it IEEE} %
\thanks{R. Gaffoglio and F. P. Andriulli are with the Department of Electronics and Telecommunications (DET), Politecnico di Torino, Torino, I-10129, Italy (e-mail: rossella.gaffoglio@polito.it, francesco.andriulli@polito.it).}
\thanks{A. Cagliero is with the Microwave Department of IMT Atlantique – Institut Mines-T\'{e}l\'{e}com, Brest, F-29238, France (e-mail: andrea.cagliero@imt-atlantique.fr).}
\thanks{G. Vecchi is with Antenna and EMC Lab (LACE), Department of Electronics and Telecommunications (DET), Politecnico di Torino, I-10129 Torino, Italy (e-mail: giuseppe.vecchi@polito.it)}
\thanks{{\textbf{2169-3536 (c) 2017 IEEE. Personal use is permitted, but republication$/$redistribution requires IEEE permission. See}}
http:$//$www.ieee.org$/$publications$\_$standards$/$publications$/$rights$/$index.html {\textbf{for more information.}}}}

\maketitle

\selectlanguage{english}

\begin{abstract}

Several recent contributions have envisioned the possibility of increasing currently exploitable maximum channel capacity of a free space link, both at optical and radio frequencies, by using vortex waves, i.e. carrying Orbital Angular Momentum (OAM). Our objective is to disprove these claims by showing that they are in contradiction with very fundamental properties of Maxwellian fields. We demonstrate that the Degrees of Freedom (DoF) of the field cannot be increased by the helical phase structure of electromagnetic vortex waves beyond what can be done without invoking this property. We also show that the often-advocated over-quadratic power decay of OAM beams with distance does not play any fundamental role in the determination of the channel DoF.

\end{abstract}

\selectlanguage{english}

\IEEEpeerreviewmaketitle

\section{Introduction}

\IEEEPARstart{O}{rbital} angular momentum (OAM) beams are well-known solutions to the Helmholtz equation, characterized by the presence of an optical vortex located on the propagation axis, where the intensity is zero and the phase is undefined \cite{Grier,Padgett}. In the mathematical formulation such phase singularity is expressed by a screw dislocation of the form $e^{im\varphi}$, where $\varphi$ is the azimuthal angle, while the topological charge $m$, related to the orbital angular momentum carried by the beam, determines the complexity of the helical structure of the phase fronts. In the last few years the study of the electromagnetic beams carrying OAM has generated great interest within the scientific community involving different research fields, such as nanotechnologies \cite{Grier,Padgett}, astronomy \cite{Harwit,Tamburini2011}, quantum physics \cite{Mair,Dada} and telecommunications \cite{Bozinovic,Krenn}. In particular, due to the orthogonality among vortex modes with different charge $m$, the possibility of exploiting the wave vorticity in a wireless communication context has been investigated in optics and later at the radio frequencies (RF) as a means to increase the information transfer at given frequency and polarization \cite{Gibson,Wang,Thide,Tamburini2012,Yan}.

It is relevant to note that waves carrying OAM can be detected only as a result of spatial correlation, i.e. by exploiting the finite size of the receiving device (antenna, aperture). Conversely, as well known, the ability to shape a beam depends on the size of the emitting device. This classifies OAM-based communication transmission within the class of systems exploiting spatial diversity or spatial multiplexing. Indeed, the only way to increase a communication channel capacity is to resort to independent sub-channels via spatial diversity/multiplexing. The OAM therefore seems a very good candidate to increase the channel capacity. However, several contributions \cite{Edfors,Tamagnone2012,Tamagnone2013,Zhao,Chen} have risen doubts on the practical advantages of OAM-based communications over more conventional schemes, especially with respect to power (for given channel noise characteristics), and conventional line-of-sight (LOS) multiple-input-multiple-output (MIMO) schemes for RF links. In particular, such works deal with comparisons of the OAM-based multiplexing method with MIMO \cite{Edfors} and other standard techniques \cite{Zhao,Chen}, or claim that the use of vortex modes is not necessary to encode different channels \cite{Tamagnone2012,Tamagnone2013}.

A different and more general approach is presented here: we show in fact that there is a fundamental physical reason why no advantage of OAM can be expected with respect to any other space diversity or space multiplexing technique. Also, there is a limit to the channel capacity added by these techniques that depends on the spatial extension of the emitting and receiving devices. Because of the generality of this result, an OAM-based multiplexing scheme cannot be exempt from this limit, whose validity does not depend on the nature of the modes employed in the transmission. Here, we stress this behavior by resorting to the concept of field degrees of freedom (DoF), and our results are not limited to the paraxial regime.

If one considers the total set of OAM beams - no matter how this set is defined - the upper bound to the number of independent signals that can be transmitted for a unit bandwidth is just the number of linearly independent wavefunctions necessary to represent this set (irrespective of the difficulty in practically receiving them). Thus, this number clearly identifies the number of Degrees of Freedom (DoF) of the radiated field, which in turn is directly related to channel capacity as described above.

Finally, we address here for the first time the issue of the duality between the OAM faster-than-quadratic power decay and the exponential limitation of the DoF.

\section{Simulation and Results}

The concept of field DoF is crucial to many applications of the theory of electromagnetic wave phenomena, where it is well assessed \cite{Slepian,Bucci,Miller,Piestun}. Quantitatively, the number of wavefunctions necessary to represent the scattered field everywhere in the surrounding space is bounded by the following upper limit \cite{Bucci}:
\begin{equation} \label{eq_f1}
N_{DoF}\leq \frac{4}{\pi}\:\left(\sqrt{2}k_0 a\right)^2,
\end{equation}
where $k_0=2\pi/\lambda$ is the wavenumber, being $\lambda$ the wavelength, while $a$ corresponds to the radius of the minimal ball enclosing the sources.

\begin{figure}[!t]
\centering
\includegraphics[scale=0.45]{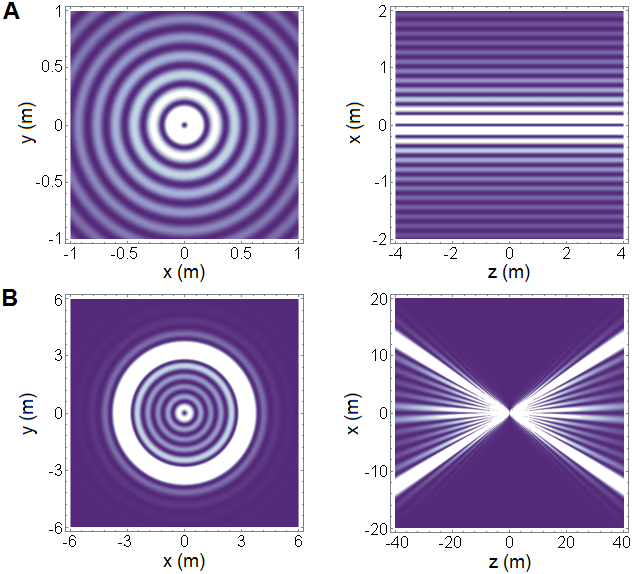}
\caption{({\textbf{A}}) Intensity profile of a $z$-directed Bessel beam with topological charge $m = 1$, wavelength $\lambda = 0.1$ m and transverse wavenumber $k_\rho=k_0\sin(\pi/10)$, displayed in the $xy$ plane at $z = 10$ m ({\it left}) and in the $zx$ plane at $y = 0$ ({\it right}). ({\textbf{B}}) Intensity profile of a Bessel beam with topological charge $m = 1$, wavelength $\lambda = 0.1$ m and transverse wavenumber $k_\rho=k_0\sin(\pi/10)$, truncated in the $z = 0$ plane by a circular aperture of radius $a = 1$ m and displayed in the $xy$ plane at $z = 10$ m ({\it left}) and in the $zx$ plane at $y = 0$ ({\it right}).}
\label{fig_Gaffo1}
\end{figure}

We will discuss the issue of vortex waves DoF employing two well-recognized embodiments of proposed OAM communications, i.e. Bessel beams generated by a continuous source distribution over an aperture (see Fig. \ref{fig_Gaffo1}), and a ring of point sources (see Fig. 4).

The first case is a relevant example of visible light wave communications and microwave aperture antennas (e.g. reflectors), the second of  RF/microwave antenna arrays which are the alternative to reflectors; also, arrays are the basis of MIMO systems.

\subsection{Truncated Bessel Beams}
We address the issue of physical limitations to OAM by first considering the Bessel beams (BB) \cite{Durnin}, that are well-recognized OAM beams. We prefer this class of beams because they are solutions of the Helmholtz equation everywhere and not only in the paraxial region, unlike other typical OAM beams, such as the set of Laguerre-Gaussian modes, often considered in previous works \cite{Zhao,Chen}. Since BB imply an infinitely extended source, in order to consider physically realizable fields, we will focus on BB produced by a finite-size aperture. Bessel Beams are defined via:
\begin{equation} \label{eq_f2}
u_m(\rho,\varphi,z)=A\:J_{|m|}(k_\rho \rho)\:e^{im\varphi}\:e^{-ik_z z},
\end{equation}
where $m$ is the topological charge of the beam, $A$ is the amplitude, $J_{|m|}$ is a Bessel function of the first kind with order $|m|$ and $k_\rho^2+k_z^2=k^2_0$. BB produced by a finite-size aperture are obtained by truncating their support and inserting (\ref{eq_f2}) in the generalized Kirchhoff diffraction integral \cite{Jackson}, which results in a closed form expression (see Appendix A for the derivation). The intensity profiles of a representative BB and of its truncated form are displayed in Fig. \ref{fig_Gaffo1}.   

\begin{figure}[!t]
\centering
\includegraphics[scale=0.53]{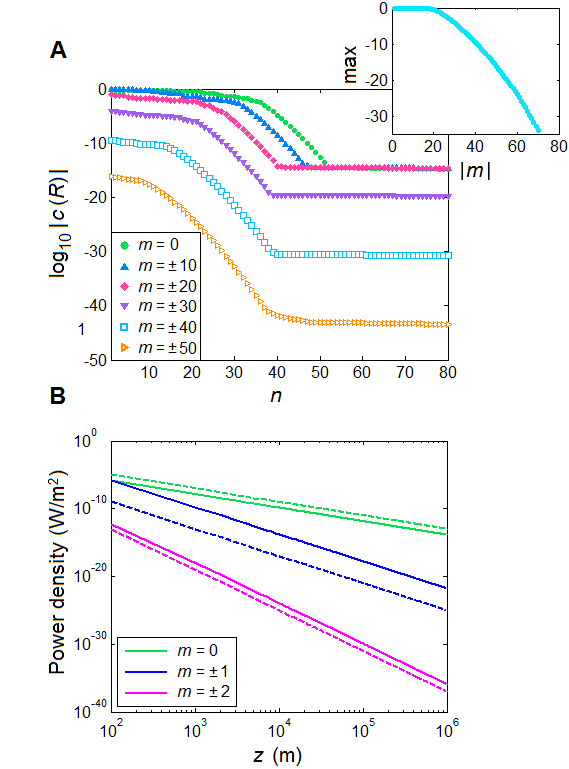}
\caption{Spherical harmonics expansion of a Bessel beam with topological charge $m$, wavelength $\lambda = 0.1$ m and transverse wavenumber $k_\rho = k_0\sin(\pi/10)$. The beam source is a circle of radius $a = 1$ m in the $z = 0$ plane and the expansion is performed over a sphere of radius $R = 5000\:\lambda$. ({\textbf{A}}) Expansion coefficients of Eq. (\ref{eq_f3}), arranged by decreasing magnitude for some values of $m$; an extended analysis of the $m$ index can be found in the inset, that shows the largest coefficient for each value of $m$. Note the exponential decay past a critical index which is independent of the beam order. ({\textbf{B}}) Power density decay along the propagation axis. The solid lines show the numerically computed power density as a function of the distance $z$; the dashed lines report the predicted polynomial power decay $z^{-2|m|-2}$.}
\label{fig_Gaffo2}
\end{figure}

\begin{figure}[!t]
\centering
\includegraphics[scale=0.5]{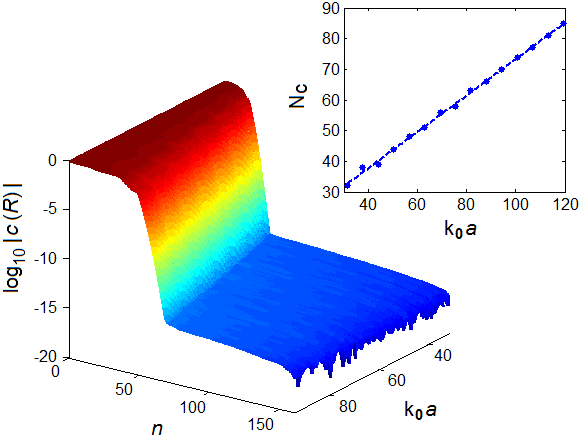}
\caption{The expansion coefficients are arranged as in Fig. \ref{fig_Gaffo2} and displayed also as a function of $k_0 a$, where $a$ is the aperture radius, for $m = 1$. The critical $N_c$ index is evaluated where the curve of the expansion coefficients becomes flat; the inset reports this as a function of the source size $k_0 a$.}
\label{fig_Gaffo3}
\end{figure}

The usual claim is that independent signals can be transported by each of the linearly independent OAM beams. The information content associated to the entire set of BB can be assessed by representing the wavefields of individual (truncated) BB over a spherical surface in terms of a multipole expansion, i.e.:
\begin{equation} \label{eq_f3}
u^{\textsc{\tiny{\it T}}}_m(r,\theta,\phi)=\sum_{\ell=0}^\infty\sum_{p=-\ell}^\ell c^{m,p}_\ell(r)\:Y^p_\ell(\theta,\phi),
\end{equation}
where $Y^p_\ell(\theta,\phi)$ are the standard spherical harmonics, being $\ell$ and $p$, with $|p|\leq\ell$, the degree and the order of the function, respectively, whereas $c^{m,p}_\ell(r)$ are the expansion coefficients. This allows to address a fundamental question: what is the number of linearly independent wavefunctions ($Y^p_\ell(\theta,\phi)$) that are necessary to represent the given wavefield with a prescribed accuracy? We have answered this question by studying the behavior of the coefficients $c^{m,p}_\ell(r)$ in (\ref{eq_f3}). This analysis is carried out in the Appendix B and the results are graphically depicted in Fig. \ref{fig_Gaffo2}A and Fig. \ref{fig_Gaffo3} (see also the supplementary Fig. \ref{fig_Gaffo5}). In these figures the expansion coefficients $c^{m,m}_\ell$, where the index $p$ has been fixed to $m$ by the presence of a Kronecker delta (see Appendix B), are displayed as a function of the spectral index $n$, which runs over $\ell\geq|m|$ according to a descending sort of the respective coefficients. As it is clear from Fig. \ref{fig_Gaffo3}, such spherical expansion coefficients have an exponential decay past a critical number $N_c \sim k_0a$, showing the same behaviour for all the values of $m$ (i.e., the vorticity of the field) included in the range $|m|\lesssim k_0a$ (Fig. \ref{fig_Gaffo2}A). Note that we have considered the entire space around the emitting source, thus providing an upper bound for the total number of estimated DoF of the source.

On the other hand, the drawback of the use of vortex waves has been typically identified with the over-quadratic decay of the associated power density \cite{Tamagnone2013,Phillips}. This decay is indeed of the type $z^{-2|m|-2}$ in the central region (Fig. \ref{fig_Gaffo2}B) for the BB (see Appendix C for more details). We stress here that the DoF limitation is instead of exponential nature and, unlike the power shortcoming, it cannot be recovered in the presence of (any) noise. This difference is further clarified by noting that the very multipole fields $\psi_{\ell m}(r,\theta,\phi)$ (see Appendix D for more discussions) show indeed an axial phase singularity (see the supplementary Fig. \ref{fig_Gaffo6}) and the same power density decay as all OAM beams - otherwise said, OAM waves ``have always been there'' in the form of spherical waves. The polynomial decay of constituent wavefunctions $\psi$ is clearly unrelated to the exponential decay of the coefficients of any wavefield representation in spherical waves past the number of DoF. 

In summary, these results state that it is not possible to increase the capacity of a communication channel by exploiting the helical phase structure of electromagnetic vortex waves beyond what can be done without invoking this property. It should also be noted that the above analysis, while explicitly carried out for BB, is completely general, and can of course be applied to the case of non-vortex waves. Indeed, any wavefield can be represented by a spherical wave expansion, whose coefficients will have a similar behavior as derived above, provided that these wavefields are a solution of the Helmholtz equation, as guaranteed by the spatial band-limitedness of these Maxwellian fields \cite{Bucci}.

\begin{figure}[!t]
\centering
\includegraphics[scale=0.52]{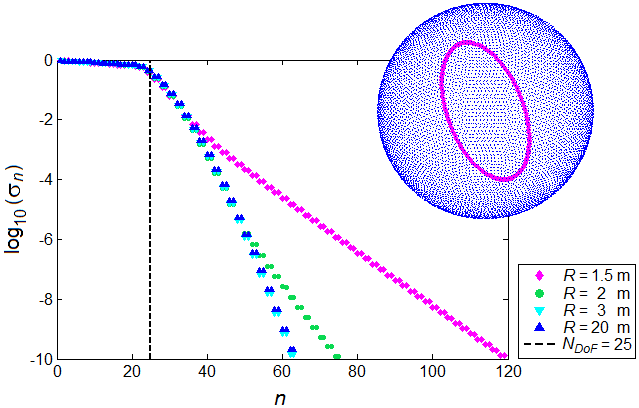}
\caption{SVD of the channel matrix $H$ for the polar ($\vartheta$) component of the electric field, for a wavelength $\lambda=0.5$ m and a ring radius $a=1$ m. In the plot $\sigma_n$ are the singular values, while $n$ is the corresponding spectral index. We consider a ring distribution of  $N = 251$ $z$-directed elementary point sources surrounded by $M = 11284$ observation points regularly arranged over a spherical surface. The change in slope of the SVD curves occurs in correspondence of the effective number of DoF predicted by the sampling theorem.}
\label{fig_Gaffo4}
\end{figure}

\subsection{Ring Distribution of Elementary Point Sources}
While the BB analysis is also common in free-space optical communications, in a Radio Frequency (RF) scenario spatial diversity is more usually associated with the use of multiple sources and receivers. To highlight the DoF importance in this scenario, we have experimented with a discrete ring distribution of linearly polarized elementary point sources in free space. The reported discussion remains unchanged, in principle, for two-dimensional arrays, where the OAM beams can be approximately reproduced by means of a (standard) array synthesis procedure. In that case however, the finite discretization of the source results in degrading the phase structure around the vortex for larger values of the topological charge. Therefore, for the sake of clarity, we have limited our discussion to the ring distribution that allows an accurate reconstruction of vortex waves with arbitrary topological charge.

For simplicity, we assume the sources to be located according to a regular spacing on the ring. Aiming at assessing an upper bound for the field DoF at a given distance, we consider receiving points regularly arranged all over a spherical surface with radius $R$ around the emitting distribution (inset of Fig. 4). With reference to common MIMO systems considerations, we can define a multiple-input and multiple-output channel via the individual links between the $n$-th element in the ring distribution and the $p$-th sampling point on the spherical observation domain. A corresponding channel matrix $H$ is introduced, whose entries $H_{pn}$ contain the electric field per unit current radiated by the $n$-th source, evaluated at the $p$-th point and tangent to the observation domain. In this context, the number of available DoF is clearly given by the numerical rank of the matrix, obtained by the singular value decomposition (SVD). The results are shown in Fig. 4, where we can see an evident change in the SVD slope that is found in agreement with the prediction on the effective number of DoF \cite{Bucci}. We now consider the channel matrix $\widetilde{H}$ in which the inputs correspond to the array synthesis of different vortex modes \cite{Thide,Nguyen} and the outputs are still related to the sampling points on the observation sphere. Simple algebra shows that such matrix can be written as the product between the original channel matrix and the discrete Fourier transform (DFT) matrix (see Appendix E for more details). Since the vortex modes are obtained by means of a linear combination of the fields of the elementary point sources, the matrices $H$ and $\widetilde{H}$ share the same spectral properties. In particular, we get for $\widetilde{H}$ the same SVD curves and thus the same effective number of DoF that was found for $H$: hence, vortex modes represent nothing but a particular basis choice in the space of the complex excitations.

\section{Conclusions}
We have discussed the field DoF as the upper bound of the number of independent communication channels (for a unit bandwidth and field polarization). About this, we have shown that the effective number of DoF of an OAM beam is bounded and only depends on the source geometry. Our results demonstrate that any system attaining the theoretically predicted limit of DoF and using an arbitrary channel discrimination strategy will not be outperformed in terms of channel capacity by a discrimination method based on vortex waves. This rules out the possibility of increasing the maximum exploitable channel capacity of a communication link with vortex waves. Our findings do not conflict with the utilization of OAM in quantum encryption \cite{Mair,Vaziri,Fickler}. However, even in this case, the propagation link segment of a quantum-encrypted communication exploiting OAM will be subjected to the above-discussed limits to channel capacity per unit bandwidth and polarization.

\appendices

\section{Truncated Bessel beams}
Neglecting the harmonic time dependence $e^{i\omega t}$, a $z$-directed Bessel beam, characterized by an optical vortex of integer charge $m$ at $\rho=0$, can be expressed in the following form \cite{Durnin}:
\begin{equation} \label{eq_A1}
\medmath{u_m(\rho,\varphi,z)=A\:J_{|m|}(k_\rho \rho)\:e^{im\varphi}\:e^{-ik_z z},}
\end{equation}
where $A$ is the amplitude, $J_{|m|}$ is a Bessel function of the first kind with order $|m|$, while $k_\rho=k_0\sin\alpha$ and $k_z=k_0\cos\alpha$ are the radial and longitudinal wavenumbers, being $k_0=2\pi/\lambda$ the modulus of the wave vector and $\alpha$ the beam {\it axicon} angle. In this section we consider an ideal Bessel beam of charge $m$ truncated by a circular aperture of radius a $a\gg\lambda$, placed in the $z=0$ plane. The aperture can be thought as an Huygens source and the beam generated at an observation point $P$ of spherical coordinates $(r,\theta,\phi)$ can be evaluated using the Huygens-Fresnel integral \cite{Jackson}:
\begin{equation} \label{eq_A2}
\medmath{u^{\textsc{\tiny{\it T}}}_m(r,\theta,\phi)=\frac{i}{\lambda}\int\limits_0^{2\pi}d\varphi\int\limits_0^a\rho\: d\rho\: u_m(\rho,\varphi,0)\cos\theta\frac{\exp(-ik_0|\vec{r}_{\textsc{\tiny{\it P}}}-\vec{r}_{\textsc{\tiny{\it S}}}|)}{|\vec{r}_{\textsc{\tiny{\it P}}}-\vec{r}_{\textsc{\tiny{\it S}}}|},}
\end{equation}
where the integration covers the whole aperture area and  indicates the position of a point $S$ of coordinates $(\rho,\varphi)$ on the circular aperture. By exploiting the circular symmetry of the aperture, (\ref{eq_A2}) for $r\gg a$ acquires the following form:
\begin{align}
\nonumber \medmath{u^{\textsc{\tiny{\it T}}}_m(r,\theta,\phi) \approx} & \medmath{\frac{i}{\lambda}\int\limits_0^{2\pi}d\varphi\int\limits_0^a\rho\: d\rho\: u_m(\rho,\varphi,0)\cos\theta \:\cdot} \\
                                & \medmath{\cdot\:\frac{\exp\left\{-ik_0[r-\rho\sin\theta\cos(\varphi-\phi)]\right\}}{r}.}   \label{eq_A3}
\end{align}

The integration over the angular variable $\varphi$ can be easily performed by taking into account the reported integral representation of the Bessel function \cite{Gradshteyn}:
\begin{equation} \label{eq_A4}
\medmath{J_n(\xi)=\frac{1}{2\pi}\int\limits_{-\pi}^\pi \exp(-inx+i\xi\sin x)dx \ \ \ \mbox{for} \ n>-1.}
\end{equation}

As a result, (\ref{eq_A3}) becomes:
\begin{align}
\nonumber \medmath{u^{\textsc{\tiny{\it T}}}_m(r,\theta,\phi)\approx} & \medmath{\:A\: i^{|m|+1}\:k_0\cos\theta\:\frac{e^{-ik_0r}}{r}\:e^{im\phi}\:\cdot} \\
                                      & \medmath{\cdot\:\int\limits^a_0 \rho\:d\rho\: J_{|m|}(k_0\rho\sin\alpha)\:J_{|m|}(k_0\rho\sin\theta).} \label{eq_A5}
\end{align}
Since the integral over $\rho$ in (\ref{eq_A5}) corresponds to the so-called Lommel's integral \cite{Whittaker}, a closed form for the truncated Bessel beam $u^{\textsc{\tiny{\it T}}}_m$ can be obtained when $\theta\neq\alpha$:
\begin{align}
\nonumber \medmath{u^{\textsc{\tiny{\it T}}}_m(r,\theta,\phi)\approx} & \medmath{\:A\:i^{|m|+1}\:\cos\theta\:\frac{e^{-ik_0r}}{r}\:e^{im\phi}\:\cdot} \\
\nonumber & \medmath{\cdot\left[\sin\theta\:J_{|m|-1}(k_0a\sin\theta)\: J_{|m|}(k_0 a\sin\alpha)+\right.} \\
\nonumber & \medmath{\left.-\sin\alpha\:J_{|m|-1}(k_0a\sin\alpha)\:J_{|m|}(k_0 a\sin\theta)\right]\cdot} \\
				  & \medmath{\cdot\frac{a}{\sin^2\alpha-\sin^2\theta},} \label{eq_A6}
\end{align}
and also when $\theta=\alpha$:
\begin{align}
\nonumber \medmath{u^{\textsc{\tiny{\it T}}}_m(r,\theta,\phi)\approx} & \medmath{\:A\:i^{|m|+1}\:k_0\:\cos\alpha\:\frac{e^{-ik_0r}}{r}\:e^{im\phi}\:\frac{a^2}{2}\:\cdot} \\
\nonumber & \medmath{\cdot\left\{\left[J_{|m|}(k_0 a\sin\alpha)\right]^2+\right.} \\
& \medmath{\left.-J_{|m|-1}(k_0a \sin\alpha)\:J_{|m|+1}(k_0a\sin\alpha)\right\}.} \label{eq_A7}
\end{align}

The so derived truncated Bessel beams contain the information about the size of the generating circular aperture. However, it is important to emphasize that the aperture radius $a$ imposes a constraint on the topological charge $m$ which limits to a finite number the Bessel beams with different $m$ that can successfully propagate without giving rise to evanescent contributions \cite{Zambrini} (see the inset of Fig. \ref{fig_Gaffo2}A).

\section{Bessel beams multipole expansion}
The number of DoF associated to a truncated Bessel beam of charge $m$ can be identified with the minimum number of orthogonal wavefunctions necessary to provide an accurate field description in a given domain. The choice to consider as the observation manifold a sphere of radius $R\gg a$ around the circular source makes the spherical harmonics $Y^p_\ell(\theta,\phi)$ the natural orthonormal basis for representing the resultant wave. According to these considerations, we perform the spherical harmonics expansion of a truncated Bessel beam with charge $m$:
\begin{equation} \label{eq_B1}
\medmath{u^{\textsc{\tiny{\it T}}}_m(R,\theta,\phi)=\sum_{\ell=0}^\infty \sum_{p=-\ell}^\ell\: c^{m,p}_\ell(R)\:Y^p_\ell(\theta,\phi),}
\end{equation}
where the expansion coefficients are given by the following expression: 
\begin{equation} \label{eq_B2}
\medmath{c^{m,p}_\ell(R)=\int\limits^{2\pi}_0d\phi\int\limits^\pi_0 d\theta\:\sin\theta\:u^{\textsc{\tiny{\it T}}}_m(R,\theta,\phi)\:{Y^p_\ell}^\ast(\theta,\phi),}
\end{equation}
while the spherical harmonics $Y^p_\ell(\theta,\phi)$ of degree $\ell$ and order $p$, with $|p|\leq\ell$, are defined as: 
\begin{equation} \label{eq_B3}
\medmath{Y^p_\ell(\theta,\phi)=\sqrt{\frac{(2\ell+1)}{4\pi}\:\frac{(\ell-p)!}{(\ell+p)!}}\:P^p_\ell(\cos\theta)\:e^{ip\phi},}
\end{equation}
being $P^p_\ell(\cos\theta)$ the associated Legendre polynomials.

The integral over $\phi$ in (\ref{eq_B2}) yields $2\pi\delta_{mp}$, where the Kronecker delta arises from the presence of the exponential terms in expressions (\ref{eq_A6}) and (\ref{eq_B3}). Hence, the spherical harmonics expansion of $u^{\textsc{\tiny{\it T}}}_m$ becomes:
\begin{equation} \label{eq_B4}
\medmath{u^{\textsc{\tiny{\it T}}}_m(R,\theta,\phi)=\sum_{\ell\geq|m|}^\infty\: c^{m,m}_\ell(R)\:Y^m_\ell(\theta,\phi).}
\end{equation}

\begin{figure}[!t]
\centering
\includegraphics[scale=0.52]{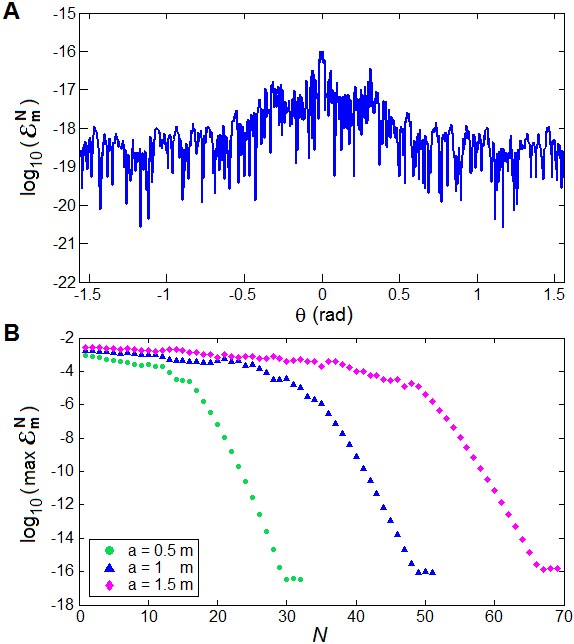}
\caption{({\textbf{A}}) Absolute error $\varepsilon^{\textsc{\tiny{\it N}}}_m(R,\theta)$ as a function of the angle $\theta$, for $m = 1$, $\lambda = 0.1$ m, $a = 1$ m, $R = 500$ m and $N$ equal to the number of expansion coefficients above the exponential decay. ({\textbf{B}}) Maximum value of the error $\varepsilon^{\textsc{\tiny{\it N}}}_m(R,\theta)$ with respect to the angle $\theta$ as a function of $N$, for $m = 1$, $\lambda = 0.1$ m, $R = 500$ m and three different values of the aperture radius $a$.}
\label{fig_Gaffo5}
\end{figure}

The expansion coefficients, sorted in $\ell$ according to the decreasing value of their modulus, show the same behaviour when studied as a function of $m$ (Fig. \ref{fig_Gaffo2}A). The number $N_c$ of coefficients $c^{m,m}_\ell(R)$ which are not exponentially suppressed (Fig. \ref{fig_Gaffo3}) indicates how many spherical harmonics are needed to represent the radiated field over the considered spherical manifold, at fixed $m$. The error performed in such reconstruction by considering the $N$ greatest expansion coefficients can be written as:
\begin{equation} \label{eq_B5}
\medmath{\varepsilon^{\textsc{\tiny{\it N}}}_m(R,\theta)=\left\|u^{\textsc{\tiny{\it T}}}_m(R,\theta,\phi)-\sum^N_{n=1}c^{m,m}_n(R)\:Y^m_n(\theta,\phi)\right\|,}
\end{equation}
where the index $n$ varies on the set of $N$ values of the spherical harmonic degree $\ell$ sorted in descending order according to the modulus of the respective expansion coefficients $c^{m,m}_\ell(R)$. The absolute error defined in (\ref{eq_B5}) slightly varies with the angle $\theta$ (Fig. \ref{fig_Gaffo5}A), while does not depend on the azimuthal angle $\phi$. By classifying as relevant the coefficients above the exponential fall, we explicitly verified that the truncated Bessel beams are actually reproduced within a small error, which is found to be lower than $10^{-16}$ for the $m = 1$ case (Fig. \ref{fig_Gaffo5}). 

In conclusion, the behaviour of the expansion coefficients is the same for all values of $m$ and only Bessel beams with $|m|\lesssim k_0a$ can propagate through the aperture. Hence, our results clearly indicate that the number of DoF is bounded and only related to the source geometry, even if the radiation is emitted by means of vortex waves.

\section{Paraxial $z$-decay}
Let us now consider the far-field evolution of the above derived truncated Bessel beams with respect to the propagation distance. The behavior of the modulus of (\ref{eq_A6}) in the paraxial region, i.e. a small transverse region around the beam axis at great distances $z$ from the aperture, can be obtained by introducing the following approximations:
\begin{equation} \label{eq_C1}
\medmath{\sin\left(\theta\right)\approx\theta, \quad \tan\left(\theta\right)\approx\theta, \quad \mathrm{and} \quad \cos\theta\approx 1.}
\end{equation}

According to (\ref{eq_C1}), $z=r\cos\theta\sim r$ and the modulus of (\ref{eq_A6}) becomes:
\begin{align}
\nonumber \medmath{\left|u^{\textsc{\tiny{\it T}}}_m(r,\theta,\phi)\right|\approx} & \medmath{\: A \: \frac{1}{z} \frac{a}{\sin^2\alpha}\left|\theta J_{|m|-1}(k_0 a\theta) \: J_{|m|}(k_0 a\sin\alpha)+\right.} \\
& \medmath{ \left. -\sin\alpha \: J_{|m|-1}(k_0 a\sin\alpha) \: J_{|m|}(k_0 a\theta) \right|},\label{eq_C2}
\end{align}
where $\theta\sim z^{-1}$. By taking into account the asymptotic forms of the Bessel functions for small arguments \cite{Abramowitz}:
\begin{equation} \label{eq_C3}
\medmath{J_{n}(x) \sim \frac{1}{\Gamma\left(n+1\right)}\left(\frac{x}{2}\right)^n \quad \mathrm{for} \quad n\geq0 \quad \mathrm{and} \quad x \approx 0},
\end{equation}
\begin{equation} \label{eq_C4}
\medmath{J_{n}(x) \sim \frac{\left(-1\right)^n}{\Gamma\left(\left|n\right|+1\right)}\left(\frac{x}{2}\right)^{\left|n\right|} \quad \mathrm{for} \quad n<0 \quad \mathrm{and} \quad x \approx 0},
\end{equation}
the paraxial behavior of the considered truncated Bessel beams with respect to the distance $z$ directly follows:
\begin{equation} \label{eq_C5}
\medmath{\left| u^{\textsc{\tiny{\it T}}}_m(r,\theta,\phi) \right| \sim z^{-1} \cdot {\left(\theta \cdot \theta^{\left|\left|m\right|-1\right|} + \theta^{\left|m\right|}\right)} \sim z^{-\left|m\right|-1}},
\end{equation}
leading to the polynomial power decay $z^{-2\left|m\right|-2}$ (see Fig. \ref{fig_Gaffo2}B).

\begin{figure}[!t]
\centering
\includegraphics[scale=0.398]{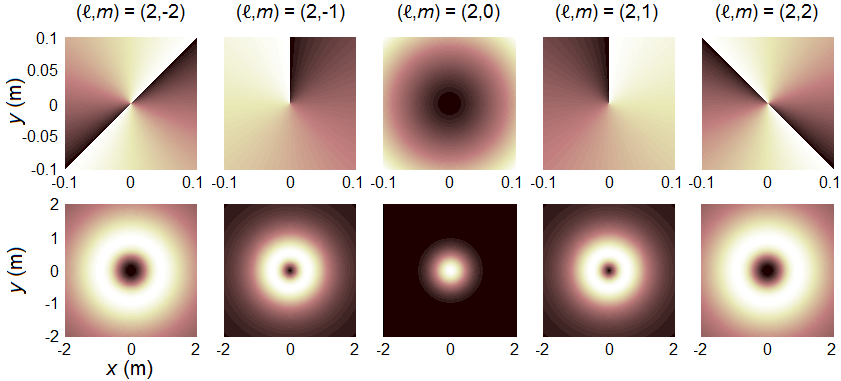}
\caption{Phase ({\it upper row}) and intensity ({\it lower row}) profiles of the multipole fields (\ref{eq_D3}) for $\ell=2$, $|m|\leq\ell$ and $k_0=1$ m$^{-1}$, displayed in the $xy$ plane at $z = 1$ m.}
\label{fig_Gaffo6}
\end{figure}

\section{Paraxial Expansion of the Scalar Multipole Fields}
Let us consider the Helmholtz equation for a scalar beam $\psi$ which is a function of the spatial coordinates:
\begin{equation} \label{eq_D1}
\medmath{\nabla^2\psi+k_0^2\psi=0},
\end{equation}
where $\nabla^2$ is the Laplace operator, $k_0=2\pi/\lambda$ is the wavenumber and $\lambda$ the wavelength. If we move to a spherical coordinate system, (\ref{eq_D1}) becomes: 
\begin{equation} \label{eq_D2}
\medmath{\frac{1}{r}\frac{\partial^2\left(r\psi\right)}{\partial r^2}+\frac{1}{r^2\sin\theta}\frac{\partial}{\partial\theta}\left(\sin\theta\frac{\partial\psi}{\partial\theta}\right)+\frac{1}{r^2\sin^2\theta}\frac{\partial^2\psi}{\partial\phi^2}+k_0^2\psi=0},
\end{equation}
and its general solution can be written in terms of the following multipole fields:
\begin{equation} \label{eq_D3}
\medmath{\psi_{\ell m}\left(r,\theta,\phi\right)=b_{\ell}\left(k_0 r\right)Y_{\ell}^m\left(\theta,\phi\right)},
\end{equation}
being $b_{\ell}\left(k_0 r\right)$ a linear combination of the spherical Bessel $j_{\ell}\left(k_0 r\right)$ and $y_{\ell}\left(k_0 r\right)$ functions and $Y_{\ell}^m\left(\theta,\phi\right)$ the spherical harmonics. Fig. \ref{fig_Gaffo6} shows the phase and the intensity profiles of (\ref{eq_D3}) for $\ell=2$. Making use of the transformation from spherical to cylindrical coordinates and taking into account the paraxial limit $z\gg \rho$ such that $\cos\theta\approx1$ and $\sin\theta \approx \theta \approx \rho / z$, we get:

\begin{align} 
\nonumber \medmath{\psi_{\ell m}\left(\rho,\varphi,z\right)\sim} & \medmath{\frac{\left(\pm i\right)^{-\ell-1}}{k_0z}\sqrt{\frac{\left(2\ell+1\right)}{4\pi}\frac{\left(\ell-m\right)!}{\left(\ell+m\right)!}}} \cdot \\
& \cdot \medmath{P_{\ell}^m\left(\cos\frac{\rho}{z}\right)\exp \left(\pm i k_0 z + i m \varphi\right)},\label{eq_D4}
\end{align}
where $P_{\ell}^m$ represent the associated Legendre polynomials, the definition $b_{\ell}\left(k_0 z\right)=j_{\ell}\left(k_0 z\right)\pm i y_{\ell}\left(k_0 z\right)$ has been introduced and the following asymptotic expression has been considered \cite{Jackson}:

\begin{equation} \label{eq_D5}
\medmath{b_{\ell}\left(k_0 r\right)\sim \frac{\left(\pm i\right)^{-\ell-1}}{k_0z} \exp \left(\pm i k_0 z\right)}.
\end{equation}

In order to guess the asymptotic behaviour of the multipole fields in the paraxial region, we must first provide an estimation to $P_{\ell}^m\left(\cos\theta\right)$ for small $\theta$. This can be done starting from the general Legendre equation:

\begin{equation} \label{eq_D6}
\medmath{\left(1-\xi^2\right)\frac{d^2 P_{\ell}^m\left(\xi\right)}{d\xi^2}-2\xi \frac{d P_{\ell}^m\left(\xi\right)}{d\xi}+\left[\ell\left(\ell+1\right)-\frac{m^2}{1-\xi^2}\right]P_{\ell}^m\left(\xi\right)=0}
\end{equation}
and considering the change of variable $\xi=\cos\theta$. If we then perform the paraxial limit of the resulting equation, we get:

\begin{equation} \label{eq_D7}
\medmath{\left\{\theta^2\frac{d^2}{d\theta^2}+\theta\frac{d}{d\theta}+\left[\ell\left(\ell+1\right)\theta^2-m^2\right]\right\}P_{\ell}^m\left(\cos\theta\right)=0},
\end{equation}
which is the Bessel equation in the variable $\sqrt{\ell\left(\ell+1\right)}\theta$. Since $P_{\ell}^m\left(\cos\theta\right)$ is not singular in $\theta=0$, we infer the following asymptotic relation:
\begin{equation} \label{eq_D8}
\medmath{P_{\ell}^m\left(\cos\theta\right)\sim J_{\left|m\right|}\left(\sqrt{\ell\left(\ell+1\right)}\theta\right)\sim \theta^{\left|m\right|}\sim\left(\frac{\rho}{z}\right)^{\left|m\right|}},
\end{equation}
where $J_{\nu}$ represents the Bessel function of the first kind, whose asymptotic expansion is provided by (\ref{eq_C3}) and (\ref{eq_C4}), and all proportionality constants have been neglected for brevity. Lastly, taking into account (\ref{eq_D4}) and (\ref{eq_D8}), we are able to express the sought-for paraxial limit of the multipole fields:

\begin{equation} \label{eq_D9}
\medmath{\lim_{z\rightarrow\infty} \psi_{\ell m}\left(\rho,\varphi,z\right)\propto\frac{\rho^{\left|m\right|}}{z^{\left|m\right|+1}}\exp \left(\pm i k_0 z +i m \varphi \right)}.
\end{equation}

Eq. (\ref{eq_D9}) tells us that, interpreting the function $\psi_{\ell m}$ in terms of a cylindrical beam and analyzing its paraxial contribution, multipole fields can be seen as vortex modes characterized by the usual polynomial power decay $z^{-2\left|m\right|-2}$ in the central region of the field intensity profile. We emphasize that such polynomial power decay should not be confused with the exponential decay of the expansion coefficients, which enables the estimation of the number of expected DoF.

\section{Ring distribution of elementary point sources}
In this section we provide all the necessary details relative to the singular value decomposition (SVD) of the channel matrix for a ring distribution of $N$ $z$-directed elementary point sources. For the sake of clarity, we report the analytic expression of the electric field radiated by an element $n$ of the distribution, evaluated at an arbitrary point $\vec{r}=(r\sin\theta\cos\phi,r\sin\theta\sin\phi,r\cos\theta)$ in the space:
\begin{equation} \label{eq_E1}
\medmath{\vec{E}_n(r,\theta,\phi)=V_0\xi_n\:\frac{\exp\left(-ik_0|\vec{r}-\vec{r}_n|\right)}{4\pi|\vec{r}-\vec{r}_n|}\:\sin\theta_n\:\hat{\theta}_n,}
\end{equation}
where $V_0\xi_n$ represents a suitable voltage coefficient and $\vec{r}_n$ corresponds to the displacement of the element with respect to the origin of the Cartesian coordinate system. Moreover:
\begin{equation} \label{eq_E2}
\medmath{\hat{\theta}_n=\left(\cos\theta_n\cos\phi_n,\cos\theta_n\sin\phi_n,-\sin\theta_n\right)}
\end{equation}
and $\theta_n$, $\phi_n$ are the polar coordinates identified by $(\vec{r}-\vec{r}_n)$ in the source element's reference frame. Being interested in a circular distribution composed by $N$ equispaced sources, $\vec{r}_n=a(0,\cos\varphi_n,\sin\varphi_n)$, where $a$ is the radius of the ring and $\varphi_n=2\pi(n-1)/N$  describes the azimuthal position of the $n$-th source. The global electric field of the ring distribution is then simply given by:
\begin{equation} \label{eq_E3}
\medmath{\vec{E}(r,\theta,\phi)=\sum_{n=1}^N\:\vec{E}_n(r,\theta,\phi)}.
\end{equation}
Let's now consider a spherical observation surface with radius $R$, placed around the ring distribution. We choose $M$ sampling points regularly arranged over the surface at the positions $\vec{r}_p=(R\sin\theta_p\cos\phi_p,R\sin\theta_p\sin\phi_p,R\cos\theta_p)$, where the $p$ index runs from 1 to $M$. The channel matrix which relates the sets of $N$ complex source excitations with the corresponding $M$ electric field values tangent to the observation manifold in each sampling point is given by:
\begin{equation} \label{eq_E4}
\medmath{\overrightarrow{H}_{pn}=V_0\:\frac{\exp\left(-ik_0|\vec{r}_p-\vec{r}_n|\right)}{4\pi|\vec{r}_p-\vec{r}_n|}\:\sin\theta_n\hat{\theta}_n\cdot\left(\hat{\theta}_p\hat{\theta}_p+\hat{\phi}_p\hat{\phi}_p\right),}
\end{equation}
where:
\begin{equation} \label{eq_E5}
\medmath{|\vec{r}_p-\vec{r}_n|=\sqrt{R^2+a^2-2aR(\sin\theta_p\sin\phi_p\cos\varphi_n+\cos\theta_p\sin\varphi_n)}}
\end{equation}
and the dyadic form $\left(\hat{\theta}_p\hat{\theta}_p+\hat{\phi}_p\hat{\phi}_p\right)$ represents a projector on the tangent plane to the sphere. Explicitly:
\begin{align}
\medmath{\nonumber \hat{\theta}_n\cdot\left(\hat{\theta}_p\hat{\theta}_p+\hat{\phi}_p\hat{\phi}_p\right)=} & \medmath{\left[\cos\theta_n\cos\theta_p\cos(\phi_n-\phi_p)+\sin\theta_n\sin\theta_p\right]\hat{\theta}_p+} \\
& \medmath{+ \cos\theta_n\sin(\phi_n-\phi_p)\hat{\phi}_p.}  \label{eq_E6}
\end{align}

By means of the SVD procedure, we write the channel matrix in (\ref{eq_E6}) in terms of the product $H=U\Sigma V^\dagger$, where $U$ and $V$ are unitary square matrices and $\Sigma$ represents a diagonal rectangular matrix whose entries correspond to the non null singular values of $H$ sorted in decreasing order. The singular value decomposition of the channel matrix in Eq. (\ref{eq_E6}) enables one to get direct access to its spectral content and thus to extrapolate the effective number of DoF.

\begin{figure}[!t]
\centering
\includegraphics[scale=0.43]{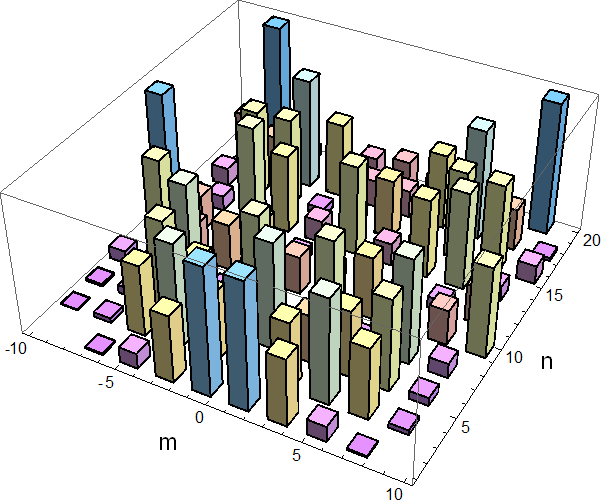}
\caption{The OAM content ($m$ index) of the channel matrix (\ref{eq_E4}) spectrum is displayed via probability histogram for some of the first right singular vectors (labelled by the spectral index $n$). Due to the directivity of the considered source elements, spurious vortex contributions naturally emerge in the channel spectrum.}
\label{fig_Gaffo7}
\end{figure}

In the analysis reported above, the columns of the channel matrix $H$ correspond to excitations with all but one zero coefficients, i.e. each element transmitting as standalone. Now, we define a new channel matrix $\widetilde{H}$ in which the multiple inputs are coded as vortex modes instead of single elementary sources. In the simple case of a ring distribution of $N$ radiating elements, the $n$-th source excitation coefficient for the vortex mode with azimuthal index $m$ is given by:
\begin{equation} \label{eq_E7}
\medmath{\xi_n^{(m)}=\frac{1}{\sqrt{N}}\:\exp(im\varphi_n).}
\end{equation}

It is an easy task to show that the vortex channel matrix $\widetilde{H}$ can be written as the product between the original matrix $H$ and the discrete Fourier transform (DFT) matrix $\Lambda$:
\begin{equation} \label{eq_E8}
\medmath{\widetilde{H}_{pj}=\sum_n H_{pn}\Lambda_{jn}.}
\end{equation}

In (\ref{eq_E8}), the DFT matrix can be written as $\Lambda_{jn}=\xi^{(m_j)}_n$, where the $j$ index runs from 1 to $N$ and the following convention has been introduced:
\begin{equation} \label{eq_E9}
\medmath{m_j}=
\medmath{
\begin{cases}
-\frac{N}{2}+j-1 \ \ \ \ \ \: N\:\mbox{even;}\\
-\frac{N-1}{2}+j-1 \ \ N\:\mbox{odd.}
\end{cases}
}
\end{equation}

It has been proven \cite{Edfors} that, in the simple case in which the elementary sources over the ring are replaced by ideal isotropic radiators, the set of excitations yielding the OAM modes directly provides the spectral basis for the corresponding channel matrix. However, elementary linearly polarized dipoles break the degenerate symmetry of the isotropic case and the interpretation of vortex modes as singular vectors of the channel matrix needs to be revisited. This circumstance can be brought to light by analyzing the spectral projection of the singular vectors of the channel matrix (\ref{eq_E4}) on the OAM basis vectors (\ref{eq_E7}), as reported in Fig. \ref{fig_Gaffo7}. It can be shown that one possible solution for restoring the lost circular symmetry is provided by the introduction of circularly polarized sources.


\begin{thebibliography}{10}
\providecommand{\url}[1]{#1}
\csname url@samestyle\endcsname
\providecommand{\newblock}{\relax}
\providecommand{\bibinfo}[2]{#2}
\providecommand{\BIBentrySTDinterwordspacing}{\spaceskip=0pt\relax}
\providecommand{\BIBentryALTinterwordstretchfactor}{4}
\providecommand{\BIBentryALTinterwordspacing}{\spaceskip=\fontdimen2\font plus
\BIBentryALTinterwordstretchfactor\fontdimen3\font minus
  \fontdimen4\font\relax}
\providecommand{\BIBforeignlanguage}[2]{{%
\expandafter\ifx\csname l@#1\endcsname\relax
\typeout{** WARNING: IEEEtran.bst: No hyphenation pattern has been}%
\typeout{** loaded for the language `#1'. Using the pattern for}%
\typeout{** the default language instead.}%
\else
\language=\csname l@#1\endcsname
\fi
#2}}
\providecommand{\BIBdecl}{\relax}
\BIBdecl

\bibitem{Grier}
D.~G. Grier, ``A revolution in optical manipulation,'' \emph{Nature}, vol. 424,
  pp. 810--816, 2003.

\bibitem{Padgett}
M.~Padgett and R.~Bowman, ``Tweezers with a twist,'' \emph{Nature Photon.},
  vol.~5, pp. 343--348, 2011.

\bibitem{Harwit}
M.~Harwit, ``Photon orbital angular momentum in astrophysics,''
  \emph{Astrophys. J.}, vol. 597, pp. 1266--1270, 2003.

\bibitem{Tamburini2011}
F.~Tamburini, B.~Thid\'{e}, G.~Molina-Terriza, and G.~Anzolin, ``Twisting of
  light around rotating black holes,'' \emph{Nature Phys.}, vol.~7, pp.
  195--197, 2011.

\bibitem{Mair}
A.~Mair, A.~Vaziri, G.~Weihs, and A.~Zeilinger, ``Entanglement of the orbital
  angular momentum states of photons,'' \emph{Nature}, vol. 412, pp. 313--316,
  2001.

\bibitem{Dada}
A.~C. Dada, J.~Leach, G.~S. Buller, M.~J. Padgett, and E.~Andersson,
  ``Experimental high-dimensional two-photon entanglement and violations of
  generalized {B}ell inequalities,'' \emph{Nature Phys.}, vol.~7, pp. 677--680,
  2011.

\bibitem{Bozinovic}
N.~Bozinovic \emph{et~al.}, ``Terabit-scale orbital angular momentum mode
  division multiplexing in fibers,'' \emph{Science}, vol. 340, pp. 1545--1548,
  2013.

\bibitem{Krenn}
M.~Krenn \emph{et~al.}, ``Communication with spatially modulated light through
  turbulent air across {V}ienna,'' \emph{New J. Phys.}, vol.~16, 2014.

\bibitem{Gibson}
G.~Gibson \emph{et~al.}, ``Free-space information transfer using light beams
  carrying orbital angular momentum,'' \emph{Opt. Express}, vol.~12, pp.
  5448--5456, 2004.

\bibitem{Wang}
J.~Wang \emph{et~al.}, ``Terabit free-space data transmission employing
  {O}rbital {A}ngular {M}omentum multiplexing,'' \emph{Nature Photon.}, vol.~6,
  pp. 488--496, 2012.

\bibitem{Thide}
B.~Thid\'{e} \emph{et~al.}, ``Utilization of photon orbital angular momentum in
  the low-frequency radio domain,'' \emph{Phys. Rev. Lett.}, vol.~99, 2007.

\bibitem{Tamburini2012}
F.~Tamburini \emph{et~al.}, ``Encoding many channels on the same frequency
  through radio vorticity: first experimental test,'' \emph{New J. Phys.},
  vol.~14, 2012.

\bibitem{Yan}
Y.~Yan \emph{et~al.}, ``High-capacity millimetre-wave communications with
  orbital angular momentum multiplexing,'' \emph{Nature Commun.}, vol.~5, 2014.

\bibitem{Edfors}
O.~Edfors and A.~J. Johansson, ``Is orbital angular momentum ({OAM}) based
  radio communication an unexploited area?'' \emph{IEEE Trans. Antennas
  Propag.}, vol.~60, pp. 1126--1131, 2012.

\bibitem{Tamagnone2012}
M.~Tamagnone, C.~Craeye, and J.~Perruisseau-Carrier, ``Comment on `{E}ncoding
  many channels on the same frequency through radio vorticity: first
  experimental test','' \emph{New J. Phys.}, vol.~14, 2012.

\bibitem{Tamagnone2013}
------, ``Comment on `{R}eply to {C}omment on {``{E}ncoding many channels on
  the same frequency through radio vorticity: first experimental test''}',''
  \emph{New J. Phys.}, vol.~15, 2013.

\bibitem{Zhao}
N.~Zhao, X.~Li, G.~Li, and J.~M. Kahn, ``Capacity limits of spatially
  multiplexed free-space communication,'' \emph{Nature Photon.}, vol.~9, pp.
  822--826, 2015.

\bibitem{Chen}
M.~Chen, K.~Dholakia, and M.~Mazilu, ``Is there an optimal basis to maximise
  optical information transfer?'' \emph{Sci. Rep.}, vol.~6, 2016.

\bibitem{Slepian}
D.~Slepian and H.~O. Pollak, ``Prolate spheroidal wave functions, {F}ourier
  analysis and uncertainty - {I},'' \emph{Bell Syst. Tech. J.}, vol.~40, pp.
  43--63, 1961.

\bibitem{Bucci}
O.~M. Bucci and G.~Franceschetti, ``On the spatial bandwidth of scattered
  fields,'' \emph{IEEE Trans. Antennas Propag.}, vol.~35, pp. 1445--1455, 1987.

\bibitem{Miller}
D.~A.~B. Miller, ``Communicating with waves between volumes: evaluating
  orthogonal spatial channels and limits on coupling strengths,'' \emph{Appl.
  Opt.}, vol.~39, pp. 1681--1699, 2000.

\bibitem{Piestun}
R.~Piestun and D.~A.~B. Miller, ``Electromagnetic degrees of freedom of an
  optical system,'' \emph{J. Opt. Soc. Am. A}, vol.~17, pp. 892--902, 2000.

\bibitem{Durnin}
J.~Durnin, ``Exact solutions for nondiffracting beams. {I}. {T}he scalar
  theory,'' \emph{J. Opt. Soc. Am. A}, vol.~4, pp. 651--654, 1987.

\bibitem{Jackson}
J.~D. Jackson, \emph{{Classical Electrodynamics}}, 3rd~ed.\hskip 1em plus 0.5em
  minus 0.4em\relax Wiley, New York, 1999.

\bibitem{Phillips}
R.~L. Phillips and L.~C. Andrews, ``Spot size and divergence for {L}aguerre
  {G}aussian beams of any order,'' \emph{Appl. Opt.}, vol.~22, pp. 643--644,
  1983.

\bibitem{Nguyen}
D.~K. {N}guyen \emph{et~al.}, ``Antenna gain and link budget for waves carrying
  orbital angular momentum,'' \emph{Radio Sci.}, vol.~50, pp. 1165--1175, 2015.

\bibitem{Vaziri}
A.~Vaziri, G.~Weihs, and A.~Zeilinger, ``Superpositions of the orbital angular
  momentum for applications in quantum experiments,'' \emph{J. Opt. B: Quant.
  Semiclass. Opt.}, vol.~4, pp. S47--S51, 2002.

\bibitem{Fickler}
R.~Fickler \emph{et~al.}, ``Quantum entanglement of high angular momenta,''
  \emph{Science}, vol. 338, pp. 640--643, 2012.

\bibitem{Gradshteyn}
I.~S. Gradshteyn and I.~M. Ryzhik, \emph{{Table of Integrals, Series and
  Products}}, 7th~ed.\hskip 1em plus 0.5em minus 0.4em\relax Elsevier, New
  York, 2007.

\bibitem{Whittaker}
E.~T. Whittaker and G.~N. Watson, \emph{{A Course of Modern Analysis}},
  2nd~ed.\hskip 1em plus 0.5em minus 0.4em\relax Cambridge Univ. Press,
  Cambridge, 1915.

\bibitem{Zambrini}
R.~Zambrini, L.~C. Thomson, S.~M. Barnett, and M.~Padgett, ``Momentum paradox
  in a vortex core,'' \emph{J. Modern Opt.}, vol.~52, pp. 1135--1144, 2005.

\bibitem{Abramowitz}
M.~Abramowitz and I.~A. Stegun, \emph{{Handbook of Mathematical Func\-tions,
  with Formulas, Graphs, and Mathematical Tables}}, 10th~ed.\hskip 1em plus
  0.5em minus 0.4em\relax National Bureau of Standards, Washington, 1972.

\end{thebibliography}
\end{document}